\documentclass[twocolumn,showpacs,preprintnumbers,amsmath,amssymb,aps]{revtex4}

\usepackage{graphicx}
\usepackage{dcolumn}
\usepackage{bm}

\newcommand{\beq}{\begin{equation}}
\newcommand{\eeq}{\end{equation}}
\newcommand{\bea}{\begin{eqnarray}}
\newcommand{\eea}{\end{eqnarray}}
\newcommand{\bean}{\begin{eqnarray*}}
\newcommand{\eean}{\end{eqnarray*}}
\newcommand{\pder}[2]{\frac{\partial {#1}}{\partial {#2}}}

\newcommand{\dam}{{\mathrm{Da }}}
\newcommand{\peclet}{{\mathrm{Pe }}}

\begin{document}

\preprint{}

\title{Persistent localized states for a chaotically mixed bistable reaction}

\author{Stephen M. Cox}
 \email{stephen.cox@nottingham.ac.uk}
 \homepage{http://www.maths.nottingham.ac.uk/personal/smc}
\affiliation{School of Mathematical Sciences,
University of Nottingham, University Park, Nottingham NG7 2RD, United Kingdom}

\date{\today}

\begin{abstract}

We describe the evolution of a bistable chemical reaction in a closed
two-dimensional chaotic laminar flow, from a localized initial
disturbance. When the fluid mixing is sufficiently slow, the disturbance
may spread and eventually occupy the entire fluid domain. By contrast,
rapid mixing tends to dilute the initial state and so extinguish the
disturbance. Such a dichotomy is well known. However, we report here a
hitherto apparently unremarked intermediate case, a persistent highly
localized disturbance. Such a localized state arises when the
Damk\"ohler number is great enough to sustain a ``hotspot'', but not so
great as to lead to global spread. We show that such a disturbance is
located in the neighborhood of an unstable periodic orbit of the flow,
and we describe some limited aspects of its behavior using a reduced,
lamellar model.

\end{abstract}

\maketitle

\section{\label{sec:intro}Introduction}

The progress of chemical or biological processes mixed by fluid flows is 
of great environmental and industrial significance, in a wide variety of 
contexts (see~\cite{Epstein95,Abraham98,Edouard96,Meron92} for some broad 
applications; in this paper we shall, for brevity, mostly use the language 
of chemical processes in our discussions). The underlying processes often 
possess multiple possible long-time states, characterized by whether some 
reaction proceeds or dies out. The ultimate fate of a localized initial 
disturbance can depend subtly on the combined influences of advection by 
the fluid, diffusion of the reactants, and the reaction kinetics.

This paper concerns the influence of chaotic, laminar fluid mixing on the 
course of a bistable chemical reaction in a closed flow domain. In a 
spatially homogeneous environment, such a reaction tends to one of two 
steady states (which we describe below as the ``quenched'' and ``excited'' 
states) according to whether or not the initial reactant concentrations 
exceed some threshold.  However, the large-time behavior of the reaction 
from a spatially localized disturbance subject to fluid mixing is a more 
subtle question, since the mixing has two competing effects on the course 
of the reaction. On the one hand, the mixing promotes the chemical 
reaction, since, according to the kinematics of the fluid motion, 
interfaces are exponentially stretched. However, the mixing also tends to 
suppress the reaction, since the repeated stretching and folding that is 
characteristic of chaotic flows generates a complicated filamentary 
pattern; if the thinning of the filaments is sufficiently rapid, it may 
ultimately quench the reaction, through diffusive dilution of the 
reactants. The loss of a stable excited state as the mixing grows in 
strength (or as the reaction rate is reduced) may be viewed as a 
saddle--node bifurcation, and has been identified in a number of reaction 
schemes~\cite{Neufeld01,Neufeld02,Neufeld02b,HG03,Kiss03b,Kiss03,Menon05}. 
For example, Neufeld et al.~\cite{Neufeld02} note that for both bistable 
and autocatalytic reactions there are two regimes: when mixing is slow, 
localized excited perturbations propagate throughout a closed domain; by 
contrast, under rapid mixing, localized perturbations decay to the uniform 
quenched state. In the propagative case, perturbations spread in the form 
of filaments.  For an excitable medium, Neufeld et al.~\cite{Neufeld02b} 
describe three scenarios for a closed flow: in the first, there is 
eventually a global excitation, which is spatially incoherent, so 
different parts of the domain are excited at different times; in the 
second, there is again a global excitation, but this time it is spatially 
coherent, and ultimately the system homogeneously decays to the unexcited 
state; in the third, the excitation enjoys no significant propagation 
before its ultimate decay. Kiss et al.~\cite{Kiss04} investigate a 
chemical reaction scheme capable of supporting both steady and oscillatory 
homogeneous states, subject to chaotic mixing. They find that ultimately 
the system reaches a spatially homogeneous state, which may be of either 
steady or oscillatory type, depending on the parameters.

We shall show here the existence of a third regime --- beyond the global
quenched and excited states --- comprising a persistent localized
disturbance.  For definiteness, we consider the model bistable reaction
scheme in which a chemical concentration $A({\bm x},t)$ satisfies, in
dimensionless form,
 \begin{equation}
 \pder{A}{t}+{\bm u}\cdot{\bm\nabla} A=
 \peclet^{-1}\nabla^2 A+\dam \, A(1-A)(A-\alpha),
 \label{eq:3d_eqn}
 \end{equation}
where $0<\alpha<1/2$~\cite{Neufeld02,Neufeld02b,Cox06}. Here
$\peclet=UL/D$ is the P\'eclet number, $\dam=kC^2/UL$ is the Damk\"ohler
number and ${\bm u}({\bm x},t)$ is the dimensionless fluid velocity
(here $U$, $L$ and $C$ are scales for the fluid velocity, lengths and
chemical concentrations, respectively; $D$ is the diffusion coefficient
and $k$ is the reaction constant). In view of the need to carry out
well resolved simulations at moderately high P\'eclet number, we shall
consider a two-dimensional problem, in which ${\bm u}$ is the
so-called ``sine flow''~\cite{Fran}. This choice of flow permits a
particularly efficient and accurate spectral numerical solution of the
problem~\cite{Adrover02,Cox04}.

Let us now discuss the behavior of (\ref{eq:3d_eqn}). First consider the 
reaction in the spatially homogeneous case. Then if the initial value of 
$A$ exceeds the threshold $\alpha$, $A\to1$ at large times. By 
contrast, if $A$ is initially less than the threshold, then $A\to0$ at 
large times. We expect a similar scenario to hold in (\ref{eq:3d_eqn}) if 
the initial state is nonhomogeneous, but the mixing is sufficiently rapid: 
an initial state is first rapidly averaged in space, then the long-time 
behavior of the system depends on whether or not, in this averaged (nearly 
spatially homogeneous) state, $A$ exceeds $\alpha$. In the opposite limit 
of no fluid motion, an initial localized disturbance spreads or becomes 
extinct depending on the initial state, but if the disturbance takes the 
form $A=A_0$ (a constant) in some finite region $\Omega$, and $A=0$ 
outside $\Omega$, then the condition for the disturbance to spread is 
roughly that $A_0$ should exceed $\alpha$. We expect a similar scenario 
for slow fluid mixing. These considerations, and the results of 
Refs.~\onlinecite{Neufeld02,Neufeld02b,Kiss03b,Kiss04}, suggest that, in a 
closed flow, the fate of a localized disturbance in (\ref{eq:3d_eqn}) is 
as follows: the system tends to a uniform state at large time, with 
$A\equiv0$ (for sufficiently rapid mixing or slow reaction or weak initial 
perturbation) or $A\equiv1$ (for sufficiently slow mixing or rapid 
reaction). Furthermore, given an initial disturbance that leads to the 
latter globally excited state for given values of $\dam$ and $\peclet$, 
one would expect that a sufficient reduction in $\dam$ would cause the 
reaction to proceed instead to the quenched state. However, we show here 
that there is an additional, intermediate scenario, in which the 
perturbation appears to remain localized for all time, spread along a 
segment of the unstable manifold of an unstable periodic orbit of the 
flow. We investigate this localized state through numerical simulation of 
(\ref{eq:3d_eqn}), in particular exploring its dependence on Damk\"ohler 
and P\'eclet numbers, and on the initial state. Finally, we shed some 
limited analytical light by consideration of a reduced, lamellar model 
that represents the balance between spreading of the hotspot due to 
reaction and diffusion, and compressive effects of the flow.

\section{Simulations} \label{sec:num-sims}

In this section we describe some numerical simulations of (\ref{eq:3d_eqn}).
Stirring of the fluid is accomplished by the ``sine flow'', which is
time-periodic with period $T$; then~\cite{Fran}
 \begin{equation} 
{\bm u}= \left\{\begin{array}{cl}(\sin2\pi y,0) & \quad
\mbox{for }m T\leq t<(m+\frac12)T, \\
                        (0,\sin2\pi x) & \quad
\mbox{for }( m+\frac12)T\leq t<(m+1)T,
           \end{array}\right. \label{eq:sfu}
 \end{equation} 
for $ m=0,1,2,\ldots$. The simulations take place in a doubly periodic
square box $0\leq x,y\leq1$. We consider the ``globally chaotic'' case
$T=1.6$, for which the Poincar\'e section appears to the eye to have no
significant regular islands. The initial condition is given by the
Gaussian concentration profile
 \begin{equation}
 A(x,y,0)=\exp \left[-100\pi((x-1/4)^2+(y-1/4)^2)/Q\right],
 \label{eq:ic}
 \end{equation}
for $0\leq x,y\leq 1$, so that the initial spatially averaged
concentration is $\langle A\rangle=0.01\times Q$ at $t=0$.
In most simulations, $Q=1$, although we also briefly consider other values
of $Q$, in the range $0<Q\leq2$.  The choice to center the initial disturbance
on the point $(0.25,0.25)$ was arbitrary, but, as we shall see, it proves
serendipitous.

We simulate (\ref{eq:3d_eqn}) subject to the flow field (\ref{eq:sfu}) and 
the initial state (\ref{eq:ic}) using a pseudospectral code, which is 
described in Ref.~\onlinecite{Cox04} (see also 
Ref.~\onlinecite{Adrover02}). The P\'eclet number in our simulations is 
limited to $10^5$, to ensure well resolved results with our available 
computational resources. Most simulations up to $\peclet=10^4$ are carried 
out using 512 Fourier modes in each of the $x$ and $y$ directions, and a 
time step of $2\times10^{-3}$; unlike the simulations reported 
elsewhere~\cite{Cox04}, which exclusively use the second-order 
time-stepping scheme ETDRK2, we mostly use here the fourth-order scheme 
ETDRK4 (see Ref.~\onlinecite{Matthews} for details of these two schemes).

In all the numerical simulations below, the sine flow is fixed to have 
period $T=1.6$, in order to avoid the appearance and disappearance of 
regular islands with varying $T$. Thus our investigation of the effects of 
fluid {\em mixing} is carried out by varying the P\'eclet number (smaller 
and larger values of this parameter corresponding, respectively, to more 
rapid and slower mixing). We also investigate the influence of the 
reaction rate and the initial amount of $A$ upon the long-time behavior of 
the system through varying the Damk\"ohler number $\dam$ and $Q$, 
respectively.

\subsection{Numerical results} \label{sec:num}

In all our simulations, we set $\alpha=0.25$. Thus for sufficiently rapid 
mixing the reaction is quenched (since the initial averaged concentration 
$\langle A\rangle=0.01Q<0.25$), and for slow enough mixing the disturbance 
spreads throughout the domain (since the peak initial concentration 
$A_{\rm max}=1>0.25$).

\begin{figure}
 
(a)
\includegraphics[width=0.7\linewidth]{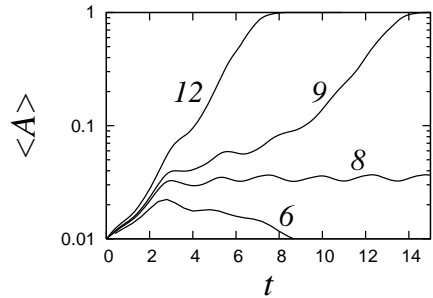}

\caption{\label{fig:Pe4} (a) Time evolution of the spatially averaged
concentration $\langle A\rangle$ for $\peclet=10^4$, from the initial
condition (\ref{eq:ic}), with $Q=1$, for various Damk\"ohler numbers,
as indicated.}

\end{figure}

We first set $\peclet=10^4$ and $Q=1$ and consider the effect of varying 
$\dam$. Fig.~\ref{fig:Pe4} shows some illustrative results. For $\dam=6$, 
the reaction is too slow to sustain itself and $A\to0$ everywhere as 
$t\to\infty$. For $\dam=12$, the reaction is fast enough to overcome the 
diluting influence of the fluid mixing; in this case $A\to1$ everywhere as 
$t\to\infty$. The choice $\dam=8$ provides an intermediate case, for which 
the reaction proceeds to neither homogeneous state as $t\to\infty$, but 
rather generates a persistent time-dependent state in which $A=0$ in most 
of the flow domain, but $A\approx1$ in a small time-dependent region. This 
region turns out, upon further examination, to lie in the vicinity of a 
period-6 orbit of the underlying flow field ${\bm u}$.  The localized 
state appears to be time-periodic, with a period of $6T$, while the 
spatial mean concentration $\langle A\rangle$ appears to have period $3T$.  
We have tracked this solution for many tens of periods, and it is robust.  
A further simulation, at $\dam=9$, at first appears to produce another 
(somewhat larger) localized ``hotspot'', but the hotspot does not approach 
a time-periodic state; instead it gradually grows, eventually filling the 
entire flow domain, so that $\langle A\rangle\to1$. 

\begin{figure}
 
\includegraphics[width=0.5\linewidth]{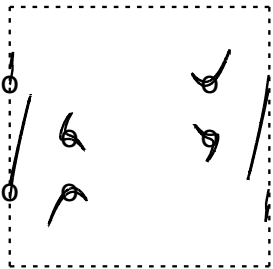}

\caption{\label{fig:PO} The circles are centered on a period-6 periodic
orbit of the sine flow (\ref{eq:sfu}), plotted at times $T$, $2T$,
\ldots, $6T$ (recall that the line $x=0$ is identified with the opposite
edge $x=1$). Also shown are the contours $A(x,y,t)=0.5$ at corresponding
times, for a hotspot found at $\dam=6$ and $\peclet=10^5$, indicating
that the hotspot is located in the vicinity of the period-6 orbit.}

\end{figure}

The localization on an unstable periodic orbit is readily understood in
qualitative terms, as follows. In the vicinity of such an orbit there
are directions of local compression and stretching. In the compressive
direction, a balance is maintained between the spreading of the
reaction--diffusion front and the compressive velocity
field~\cite{Neufeld02,Neufeld02b,Cox06}, resulting in a finite ``width''
for the hotspot. The hotspot is also stretched along the unstable
manifold of the periodic orbit; growth in its length seems to be
moderated by the inability of the reaction to sustain a thin filamentary
structure in the vigorous fluid mixing. 

We have located this unstable period-6 orbit, using the stabilization 
algorithm of Pingel, Schmelcher and Diakonos~\cite{Pingel} (see 
Fig.~\ref{fig:PO}): at times $mT$, where $m$ is an integer, the points 
$(x_m,y_m)$ on the orbit are $\cdots\to(0,0.2935)\to (0.7032,0.5) \to 
(0.7032,0.7065)\to (0,0.7065)\to (0.2968,0.5)\to 
(0.2968,0.2935)\to\cdots$. We note that, allowing for the spatial 
periodicity of the flow in both $x$ and $y$, we have 
$(x_{m+3},y_{m+3})=(1-x_m,1-y_m)$, corresponding to a rotation of $\pi$ 
radians about the center of the flow domain.  A corresponding symmetry may 
be observed in the concentration contours in Fig.~\ref{fig:PO} (following 
from the point symmetry of the velocity field ${\bm u}$ about the center 
of the flow domain), explaining why $\langle A\rangle$ has period $3T$ 
rather than $6T$. Recall that our initial condition (\ref{eq:ic}) is 
centered on the point $(0.25,0.25)$, which is close to the point 
$(0.2968,0.2935)$ on the period-6 orbit; this fortunate proximity 
underlies our finding the hotspot.

We compute that the Jacobian matrix associated with the period-6 orbit has 
eigenvalues $18.73$ and $0.053$. In the next section we shall use this 
information regarding the stretch rate close to the periodic orbit in the 
development of a reduced model for the hotspot.

\begin{figure}
 
\includegraphics[width=0.7\linewidth]{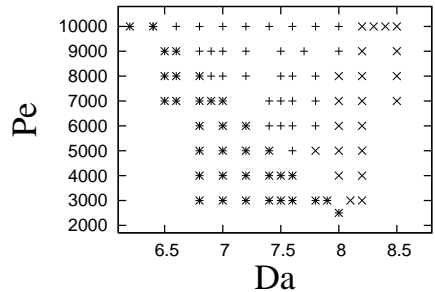}

\caption{\label{fig:par}Long-time behavior of (\ref{eq:3d_eqn}), subject 
to (\ref{eq:sfu}) and (\ref{eq:ic}) with $Q=1$, in $\dam$--$\peclet$ 
parameter space. The three long-time fates are: $\langle A\rangle\to0$ 
($*$); a localized hotspot ($+$); $\langle A\rangle\to1$ ($\times$).}

\end{figure}

Fig.~\ref{fig:par} summarizes the results of a large set of simulations, 
all at $Q=1$, for different Damk\"ohler and P\'eclet numbers. Since the 
goal of the simulations was to determine the boundaries in parameter space 
between the three different long-time behaviors, we have not attempted to 
fill in data where the behavior is (presumably) predictable. (It should be 
noted that each simulation currently takes on the order of a day or two to 
run on a desktop PC.) Gaps in the data towards the lower right-hand part 
of the figure reflect the increased length of runs required there for the 
long-time behavior to become apparent. 

Consider first the parameter dependence to the left of 
Fig.~\ref{fig:par}: if we start in the region of hotspots and decrease 
either $\dam$ or $\peclet$ then we move into the $\langle A\rangle\to0$ 
region. This is because either the reaction rate becomes too low to 
sustain the hotspot, or the initial rate of spreading of the chemical 
becomes too great for an adequate level of material to be delivered to the 
period-6 orbit in the first instance. Hence the system is quenched. To the 
right of the figure, if we start in the hotspot region and increase $\dam$ 
then we move into the $\langle A\rangle\to1$ region, because the reaction 
rate becomes too great for the hotspot to remain localized. Similarly, if 
we instead reduce $\peclet$, then again the hotspot cannot remain 
localized, due to its enhanced rate of spreading through diffusion,
and the system is globally excited. The 
parameter dependence of the system towards the lower right-hand side of 
the data in Fig.~\ref{fig:par} is rather subtle, since a reduction in 
$\peclet$ may lead either to the quenched state or to the excited state. 
Indeed, we find that our numerical simulations in this region of parameter 
space often involve a transient, slowly evolving hotspot before eventually 
selecting the ultimate quenched or excited state.

\begin{figure}
 
\includegraphics[width=0.7\linewidth]{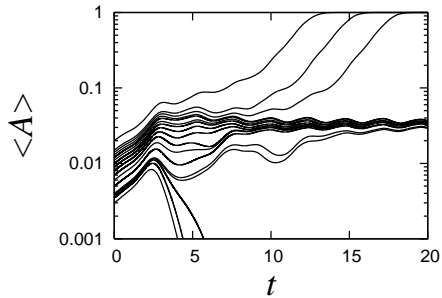}

\caption{\label{fig:size} Time evolution of $\langle A\rangle$ for
$\dam=8$ and $\peclet=10^4$, from the initial condition (\ref{eq:ic}),
for various values of $Q$. (The value of $Q$ corresponding
to each curve may be determined as 100 times the initial value of
$\langle A\rangle$.) For $Q=2,1.5,1.4$, $A\to1$ everywhere at long time; for
$Q=0.38,0.35,0.3$, $A\to0$ everywhere at long time. Intermediate values of
$Q$ lead to the (same) hotspot solution; such results are shown for $Q=0.39$
and for $Q$ from $0.4$ to $1.3$ in steps of $0.1$.}

\end{figure}

We next discuss the influence of the initial amount of $A$ upon the 
long-time behavior of the system. To this end, we have carried out 
simulations with $\peclet=10^4$ and $\dam=8$ for a range of values of $Q$. 
As Fig.~\ref{fig:size} shows, $Q\geq1.4$ leads to the globally excited 
state, $Q\leq0.38$ leads to the globally quenched state, and intermediate 
values of $Q$ lead to a hotspot at long times. Although it is not 
apparent from the figure, which focuses on relatively short-time results,
the long-time hotspot solution is the same for all of these simulations
with $Q$ in the appropriate range (this is evident in a sample of
longer simulations that we have carried out, but which
are not reported here). In other words, for these simulations 
the ultimate hotspot solution depends only on the system parameters (i.e., 
$T$, $\peclet$ and $\dam$); the role of the initial condition is merely to 
determine which of the three long-time states is selected.

\begin{figure}
 
\includegraphics[width=0.7\linewidth]{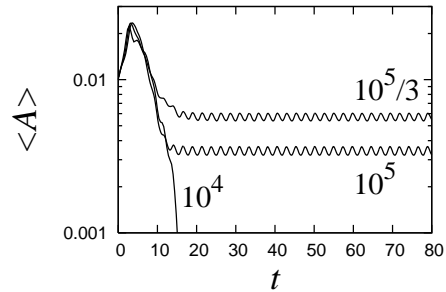}

\caption{\label{fig:k6} Time evolution of $\langle A\rangle$ for
$\dam=6$, from the initial condition (\ref{eq:ic}), for various
P\'eclet numbers, as indicated.}

\end{figure}

We now turn to the P\'eclet number dependence of the hotspot, 
characterizing the rate of fluid mixing.  In Fig.~\ref{fig:k6}, we fix the 
reaction rate $\dam=6$ and $Q=1$ and vary $\peclet$. Here we find that at 
$\peclet=10^4$, as we have described above, the relatively rapid mixing 
quenches the reaction, so that $A\to0$ everywhere. At $\peclet=10^5/3$ and 
at $\peclet=10^5$, by contrast, the mixing is slower, and we again find a 
persistent hotspot. If we accept the qualitative picture above, that the 
width of the hotspot is determined by a balance between compression and 
reaction--diffusion, while its length is determined by mixing in the bulk 
then the width should scale as $\peclet^{-1/2}$ for large $\peclet$, and 
the length should remain roughly independent of $\peclet$. Then the mean 
concentration $\langle A\rangle$ should scale as $\peclet^{-1/2}$. For the 
two hotspots indicated in Fig.~\ref{fig:k6}, this scaling is borne out, 
although admittedly over a very short range: we find, for the minimum 
value of $\langle A\rangle$ over the cycle, that $\peclet^{1/2}\langle 
A\rangle_{\mathrm{min}}=0.98$ and $1.02$ at $\peclet=10^5/3$ and $10^5$, 
respectively. (These values would be equal if the P\'eclet number scaling 
of $\langle A\rangle$ were exact.) The corresponding values of the maximum 
mean concentrations over the cycle are $\peclet^{1/2}\langle 
A\rangle_{\mathrm{max}}=1.15$ and $1.10$. If the qualitative argument 
holds, then one should expect that as $\peclet$ is increased beyond 
$10^5$, the hotspot persists, with $\langle A\rangle$ continuing to scale 
with $\peclet^{-1/2}$. Unfortunately we are unable to simulate reliably 
beyond $\peclet=10^5$, so we are unable to confirm this prediction.


\section{Lamellar model}
\label{sec:models}

There has recently
\cite{Neufeld01,Neufeld02,Neufeld02b,HG03,Kiss03b,Kiss03,%
Menon05,Cox06,Martin00,McLeod02} been significant progress in
understanding the dynamics of reaction--advection--diffusion problems
such as (\ref{eq:3d_eqn}) by analysing reduced ``filamental'' or
``lamellar'' models~\cite{r79}, which concern the simpler problem of the
evolution of the reaction in a one-dimensional setting, in our case
 \begin{equation}
\label{gen_eqn}
\frac{\partial A}{\partial t} - \lambda \xi \frac{\partial A}{\partial
\xi}= \peclet^{-1} \frac{\partial^2A}{\partial \xi^2} + \dam \,
A(1-A)(A-\alpha).
 \end{equation}
Here $\xi$ represents a coordinate in the compressive direction near an
unstable periodic orbit of the sine flow. This is the ``Lagrangian
filament slice model'', which has been thoroughly reviewed by
T\'el et al.~\cite{Tel05}, largely, but not entirely, in the
context of open rather than closed flows.

This model might be expected to provide at least a partial picture of
the behavior of the hotspot. To see this, we first note that the hotspot
is a thin structure, which appears, from the two-dimensional numerical
simulations of Sec.~\ref{sec:num}, to be approximately concentrated
around the unstable manifold of an unstable periodic orbit. Thus we
approximate the flow in the neighborhood of the orbit by its
linearization. The key balance in determining the existence or otherwise
of the hotspot is that between the compressive flow in the direction of
the stable manifold (represented by the term $-\lambda\xi A_\xi$ in
(\ref{gen_eqn})) and the tendency of the hotspot to spread as a
reaction--diffusion front (represented by the right-hand side of
(\ref{gen_eqn})), and so we reduce consideration to one-dimensional
variations along the stable manifold, and ignore variations along the
unstable manifold. This remains a complicated model, and so we take the
further gross liberty of assuming that the flow field along the stable
manifold is time-independent, with a uniform and constant rate of
compression $\lambda$. For example, for the period-6 orbit, the value of
$\lambda$ is chosen such that $\exp (-6\lambda T)$ corresponds to the
compressive eigenvalue of the Jacobian matrix of the periodic orbit
(i.e., $\lambda=(-\log0.0534)/(6T)=0.3052$). We are thus led to
(\ref{gen_eqn}) as a simplified one-dimensional model. Since nontrivial
solutions to (\ref{gen_eqn}) are localized around the origin, we may
make the further simplification that the corresponding spatial domain is
$-\infty<\xi<\infty$.

\begin{figure}
 
\includegraphics[width=0.7\linewidth]{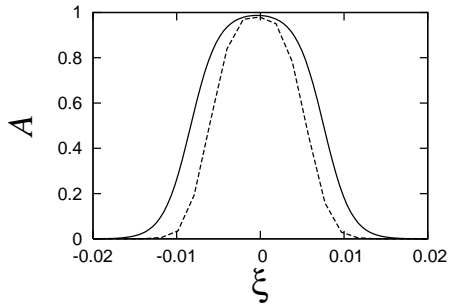}

\caption{\label{fig:cs} Cross-section through the hotspot at
$\peclet=10^5$ and $\dam=6$. Dashed line: from (\ref{eq:3d_eqn}), with
$\xi$ being a coordinate roughly normal to the hotspot. Solid line:
corresponding prediction from (\ref{gen_eqn}).}

\end{figure}

Let us now briefly recall some relevant known results for (\ref{gen_eqn}). 
An exact characterization of the behavior of a single reaction--diffusion 
front in (\ref{gen_eqn}) is possible in the absence of the advection term 
$-\lambda\xi A_\xi$~\cite{Murray,Neufeld02,Tel05}: such a front between 
$A=1$ as $\xi\to-\infty$ and $A=0$ as $\xi\to\infty$ moves with velocity 
$v=(\dam/2\peclet)^{1/2}(1-2\alpha)$. Furthermore, its profile is readily 
determined to be $A=\left\{1+\exp[\mu (\xi-vt)]\right\}^{-1}$ (up to some 
shift in time or space origin), where $\mu=(\dam\peclet/2)^{1/2}$.

In the presence of the advection term, such an exact characterization of 
solutions to (\ref{gen_eqn}) is no longer possible, although 
(large-$\dam$) asymptotic and numerical treatments are available~\cite{Cox06}.
Of particular interest in the present context is the behavior of solutions to 
(\ref{gen_eqn}) on $-\infty<\xi<\infty$ from localized initial 
disturbances, which is well 
documented~\cite{Neufeld02,Menon05,Cox06,Tel05}: if $\dam/\lambda$ is less 
than some threshold $f(\alpha)$ then all initial states decay to 
$A\equiv0$; however, if $\dam/\lambda>f(\alpha)$ then an above-threshold 
initial state generally tends to a nontrivial, stable, steady localized 
state as $t\to\infty$. The transition at $\dam/\lambda=f(\alpha)$ is a 
saddle--node bifurcation~\cite{GottwaldKramer04,Menon05}. For 
$\alpha=0.25$, as in our two-dimensional simulations, we find this 
saddle-node bifurcation to be at $f(0.25)=11.13$ (by solving 
(\ref{gen_eqn}) numerically and tracking the stable steady state as $\dam$ 
is reduced).  Since $\lambda=0.3052$, we thus conclude that no hotspots 
should be possible in the two-dimensional simulations for $\dam<3.4$. It 
is certainly the case that we have observed no hotspots for such small 
Damk\"ohler numbers; indeed, the lowest Damk\"ohler number for which we 
{\em have} observed hotspots is $6$. Given the small sample of parameter 
space provided by our simulations and the gross assumptions made in 
arguing for (\ref{gen_eqn}), this seems to represent reasonable agreement 
between this model and the full simulations. Further quantitative 
comparisons may be made by examining the structure of the hotspot itself. 
To this end, Fig.~\ref{fig:cs} shows a cross-section of the concentration 
through the hotspot, from two-dimensional simulations of (\ref{eq:3d_eqn}) 
and from the model (\ref{gen_eqn}). The general concentration profile is 
captured well by the model, although the hotspot itself is rather narrower 
than the one-dimensional model prediction. The one-dimensional model also 
predicts that only sufficiently wide Gaussian initial conditions will 
generate a hotspot, for supercritical Damk\"ohler numbers~\cite{Cox06}. 
This dependence upon the initial condition is qualitatively consistent 
with the full system, as illustrated in Fig.~\ref{fig:size}.

Of course, the one-dimensional model (\ref{gen_eqn}) has only limited 
predictive power for the full two-dimensional system. While the 
saddle--node bifurcation of the model seems to capture the demise of the 
hotspot as the Damk\"ohler number is decreased, as illustrated in 
Fig.~\ref{fig:Pe4}, there is no mechanism in the model for global 
excitation (instead the model predicts a hotspot whose width grows like 
$\dam^{1/2}$ for large $\dam$~\cite{Cox06}).  In other words, the model 
predicts behavior qualitatively consistent with the left-hand boundary of 
the hotspot region in Fig.~\ref{fig:par}, but not the right-hand boundary. 
A further shortcoming of the model is that its dependence upon the P\'eclet 
number can be removed by a straightforward scaling of $\xi$, unlike the 
full two-dimensional problem. As a consequence, the saddle--node 
bifurcation in (\ref{gen_eqn}) is independent of $\peclet$ and, for a 
given Damk\"ohler number, the stable steady-state solution simply becomes 
thinner as the P\'eclet number is increased, with width proportional to 
$\peclet^{-1/2}$. Such behavior is in contrast to the precipitate loss of 
the hotspot illustrated in Fig.~\ref{fig:k6} for $\peclet=10^4$, and the 
left-hand boundary of the hotspot region in Fig.~\ref{fig:par}.

\section{Conclusions and discussion}

For a bistable chemical reaction mixed in a chaotic laminar fluid flow, we 
have demonstrated numerically localized states that seem to persist 
indefinitely. In a sweep of parameter space, these localized states 
provide a case intermediate between global quenching and global excitation 
of the reaction, as the Damk\"ohler number is varied.  They are associated 
with unstable an periodic orbit, and their existence may partially be 
understood in terms of a simple lamellar model.

The long-time behavior of the system is strongly dependent on the size of 
the initial disturbance: too large or small and the system approaches, 
respectively, globally excited or quenched states; only for initial 
disturbances of intermediate size can a hotspot be found.

The broad influence of the Damk\"ohler number upon the long-time state of 
the system is straightforward to summarize: if we begin with parameter 
values that lead to a hotspot, then a sufficient increase or decrease in 
the Damk\"ohler number takes the system instead to the globally excited or 
quenched states.

The influence of the P\'eclet number is slightly more complicated. Again 
suppose that we start with parameter values leading to a hotspot, and 
consider the effects of reducing the P\'eclet number (i.e., enhancing 
diffusion). For lower reaction rates (i.e., smaller Damk\"ohler numbers), 
we find that the enhanced initial smearing of a localized initial 
disturbance results ultimately in a globally quenched state. When the 
reaction rate is greater, however, the nascent hotspot is able to sustain 
itself even in the presence of the enhanced diffusive spreading, but is 
unable to remain localized when the P\'eclet number is reduced 
sufficiently; thus the system approaches the globally excited state.

Since we are unable accurately to simulate P\'eclet numbers greater than 
approximately $10^5$, any large-P\'eclet number trends are inaccessible to 
our numerics. Nevertheless, we conjecture that a hotspot generally 
persists as the P\'eclet number is increased (all other parameters being 
fixed), with the amount of chemical present scaling as $\peclet^{-1/2}$. 
Of course, an exception to this suggestion would arise if one chose an 
initial condition for which diffusion was essential in transporting enough 
of an initial patch of chemical to the vicinity of the periodic orbit upon 
which the hotspot would ultimately be attached. Then it is possible that 
too large a P\'eclet number would lead to insufficient material reaching 
the intended periodic orbit, and hence to a globally quenched state.

Our discovery of these ``hotspots'' resulted from a lucky
choice of localized initial state, which happened to be centered close
to an unstable period-6 orbit. However, since the initial disturbance
lay in the chaotic region (apart from regular islands too small to see
in a Poincar\'e section of the flow) the initial state must also have
covered infinitely many other unstable periodic orbits, and we have no
explanation for why one particular orbit should have been selected as
the location of the hotspot.  It seems likely that other choices of
initial conditions and parameter regimes will lead to robust hotspots
besides the period-6 orbit evidenced here.

Although we have demonstrated hotspots for a single specific model reaction 
scheme, we expect similar hotspots to be observable in other stirred 
multistable chemical reactions (subject to sufficient luck or judgment
in choosing the initial conditions and parameter values).

Finally, we note that the existence and indefinite coherence of the
hotspots reported here is due to the periodic nature of the mixing flow
${\bm u}$. In a randomized sine flow, for example, in which the phase of
the underlying shear flows is chosen randomly at each period, we would
not expect a similar persistent phenomenon. 

\begin{acknowledgments}

The author is grateful to Georg Gottwald, Predrag Cvitanovi\'c and
Alexander Vikhansky for helpful discussions. Most of the calculations
reported here took place on the University of Nottingham High Performance
Computing grid.

\end{acknowledgments}


\end{document}